\documentclass[twocolumn,pre,floatfix]{revtex4}
\usepackage{psfrag,epsfig,amsfonts,amssymb,amsmath}
\usepackage{epstopdf}
\usepackage{dcolumn}

\newcommand{\bx}{{\bf x}}
\newcommand{\bu}{{\bf u}}
\newcommand{\bJ}{{\bf J}}
\newcommand{\bE}{{\bf E}}
\newcommand{\bn}{{\bf n}}
\newcommand{\NA}{N_{\mathrm{A}}}
\newcommand{\kB}{k_{\mathrm{B}}}

\newcommand{\seff}{\sigma_e}

\begin{document}

\title{Lateral trapping of DNA inside a voltage gated nanopore}

\author{Thomas T\"ows}
\author{Peter Reimann}
\affiliation{Fakult\"at f\"ur Physik, 
Universit\"at Bielefeld,  33615 Bielefeld, Germany}

\begin{abstract}
The translocation of a short DNA fragment through 
a nanopore is addressed when the perforated
membrane contains an embedded electrode.
Accurate numerical solutions of the
coupled Poisson, Nernst-Planck, and 
Stokes equations for a realistic, 
fully three-dimensional setup as 
well as analytical approximations 
for a simplified model are worked 
out.
By applying a suitable voltage to the membrane
electrode, the DNA can be forced to preferably
traverse the pore either along the pore axis
or at a small but finite distance from the 
pore wall.
\end{abstract}

\maketitle

\section{Introduction}
\label{s1}
The detection, analysis, and manipulation
of DNA and other macromolecules by means 
of solid-state nanopores 
\cite{dek07,hee16}
is currently a particularly
active branch of nanoscience and -technology 
\cite{nat11}.
With respect to one of the central objectives 
in this context, namely DNA sequencing, 
important breakthroughs have been achieved
in recent years,
yet a number of key issues are still
considered as not satisfactorily resolved
\cite{sch12}.
In particular, at relatively weak externally
applied driving fields, thermal noise 
effects are still too large, while at
stronger fields the DNA translocates 
through the pore too quickly to resolve
single nucleotides according to their
minimally differing current signals 
\cite{bra08}.

One promising idea to slow down the 
translocation dynamics is to integrate
an electrode into the porous membrane,
giving rise to precisely localized and/or
time-dependent forces on the DNA in
and close to the nanopore.
For example, Shankla and Aksimentiev
explored in Ref. \cite{sha14} how
the conformation and adhesion 
of single stranded DNA can be controlled
by electrically biasing a graphene 
membrane.
Here, we will specifically focus 
on embedded membrane electrodes,
i.e., they are covered by a 
thin layer of non-conducting 
material.
Such a ``nanopore capacitor'' has been
experimentally realized for the first 
time in 2006 by Gracheva et al.
to demonstrate the detectability of short 
DNA fragments \cite{gra06}.
By means of a theoretical multi-scale modeling
they furthermore demonstrated that by 
optimizing (in particular downsizing)
the system, even single nucleotides 
may become detectable.
In subsequent works they also investigated
various modifications of the original
setup by theoretical means
\cite{gra06b,sig08,mel12,jou14}.
Besides the detection, also the possibility
of steering the DNA translocation 
with the help of membrane electrodes
was theoretically predicted.

A rather similar experimental system
was realized by Yen et al in Ref. 
\cite{yen12},
reducing the DNA translocation speed
by a factor of up to 20 by applying an 
appropriate gate voltage to the membrane 
electrode,
and thus confirming earlier theoretical
predictions in Refs. \cite{lua10a,ai10,he11}.
Solid state nanopores with integrated
membrane electrodes were also successfully
fabricated by Nam et al. \cite{nam09} 
and by Jiang and Stein \cite{jia11},
while Lieber and co-workers realized 
a DNA sensor, implementing a field effect
transistor by means of a nanopore
and a narrow membrane electrode \cite{xie12}.

Systems involving more than one embedded
membrane electrodes have been addressed
experimentally in Ref. \cite{bai14}
and theoretically in Refs.
\cite{mel12,lua10,lua11}.
In particular, Stolovitzky and co-workers 
\cite{lua10,lua11} demonstrated by means
of molecular dynamics simulations a base-by-base 
ratcheting motion of single stranded DNA 
through a solid-state nanopore.
However, this model is based on several
assumptions which may be difficult to
meet in an experiment:
the applied gate voltages are very high;
the pore is assumed to be infinitely long
or to implement periodic boundary 
conditions;
the pulling force on the DNA must be 
applied via a soft spring,
which is indispensable for the
ratcheting motion.

In our present work, the main focus
is on the lateral forces, i.e. the 
force components acting upon the DNA 
during its passage through the pore
along the direction perpendicular 
to the pore axis.
In particular, we explore the possibility
to control those lateral forces by the
voltage which is applied to an embedded
membrane electrode.
Our analytical and numerical analysis
shows that osmotic pressure effects
due to the high ion concentrations
near the pore wall and the DNA play a
particularly prominent role 
in this context.
We show that by suitably choosing the gate
voltage, the DNA can be forced to
traverse the pore either along the 
pore axis, or at a small but finite 
distance from the pore wall or can 
even be forced to touch the wall.
This relatively simple way
to control the lateral motion
of the DNA opens interesting new
possibilities to optimize the
detectability and translocation 
speed of the DNA, or to render
the above mentioned ratcheting 
mechanism easier realizable
in an experiment.

The plan of the paper is as follows:
In the next section we specify 
the considered setup, the typical 
parameter values, our modeling  
in terms of the coupled Poisson, 
Nernst-Planck, and Stokes equations,
and the potential energy, from which
the lateral force derives.
In Sect. \ref{s3} we propose a simplified,
effectively one-dimensional approximation,
which can be treated analytically and
admits valuable insights into the main 
physical mechanisms at work.
In Sect. \ref{s4} an improved, 
effectively two-dimensional approximation 
is considered, which can be explored
with relatively moderate numerical effort,
and can still be quite well explained
by means of the one-dimensional model.
In Sect. \ref{s5}, the fully 
three-dimensional problem is 
numerically tackled, which in turn 
is found to often be not too far off 
the two-dimensional approximation.
The final section contains our 
conclusions and the outlook on
experimental applications.

\section{Model}
\label{s2}
\subsection{Considered setup and typical parameter values}
\label{s21}

\begin{figure}
\epsfxsize=1.05\columnwidth
\epsfbox{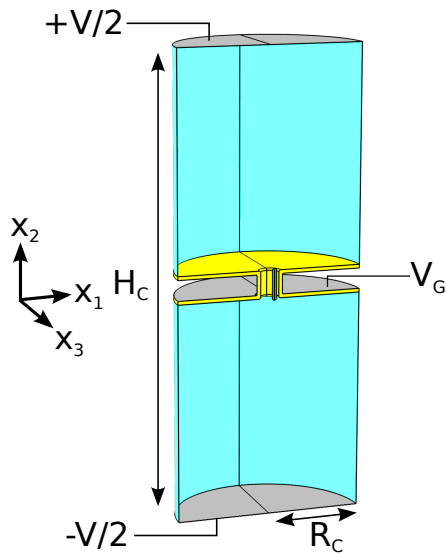} 
\caption{\label{f1}
(Color online)
Schematic cross-section through the considered setup.
A cylindrical container of height $H_c$ and 
radius $R_c$ is filled with electrolyte 
solution (blue) and is divided by a membrane. 
The membrane contains a coated 
gate electrode, a cylindrical nanopore,
and an elongated particle,
see also Fig. \ref{f2} for a 
close-up.
External voltages $V/2$, $-V/2$, and $V_G$ 
can be applied to the two electrodes
at the top and the bottom of the container, 
and to the membrane gate electrode, respectively.
The origin of the Cartesian coordinates is 
located at the pore center, but is drawn
here off the center for better visibility.
The intersection with the $x_1=0$ plane
is indicated as thin black line
to support the perspective view.
The entire sketch is drawn to scale,
adopting the standard values from 
Sect. \ref{s21} which will be used 
for the numerical solutions in 
Sect. \ref{s5}.
}
\end{figure}

\begin{figure}
\epsfxsize=0.7\columnwidth
\epsfbox{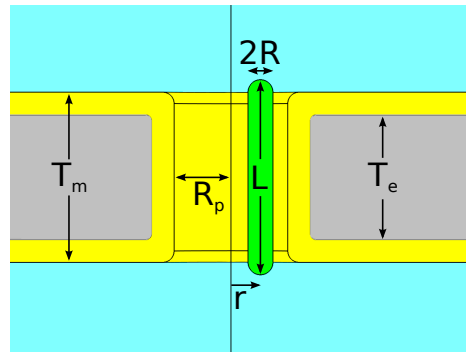} 
\caption{\label{f2}
(Color online)
Close-up of the pore region 
from Fig. \ref{f1} (again drawn to scale) 
with pore radius $R_p$, gate electrode 
(gray) of thickness $T_e$, coating (yellow), 
and total membrane thickness $T_m$. 
The particle (green) of radius $R$, 
length $L$, and distance $r$ from the pore axis 
(black line) may model, e.g., a DNA segment or an 
elongated nanoparticle.
}
\end{figure}

As depicted in Figs. \ref{f1} and \ref{f2},
we consider a system consisting of a cylindrical 
container of height $H_c$ and radius $R_c$, 
which is divided into two compartments
by a membrane of thickness $T_m$. 
The usual values adopted in our 
calculations below will be
$H_c=250\,$nm, $R_c=50\,$nm, and $T_m=15\,$nm.
The two compartments are 
connected by a nanopore and are filled with 
an electrolyte solution with a bulk 
ion concentration of $c_0$.
To keep things as simple as possible,
we restrict ourselves to cylindrical
pores (radius $R_p$).
Other pore shapes lead to similar results 
but the details become considerably more
complicated.
Quantitatively, we will focus on
$R_p=5\,$nm, and on KCl solutions with
$c_0$ in the range of $10-100\,$mM \cite{dor09,dek07,cha04,spi11,gal14}.
Accordingly, we assume that there are only two 
relevant charged species ($i=1,2$) inside 
the solution, whose diffusion coefficients 
$D_i$ take the known values
$D_1\simeq D_2\simeq 2\cdot10^{-9}$m$^2$/s 
for K$^+$ and Cl$^-$ ions in diluted
KCl solutions \cite{pro03}.
Here and in the following, we tacitly
focus on systems at room temperature.

The membrane is assumed to be a metal 
electrode of thickness $T_e$, coated by a 
dielectric layer of uniform thickness
$(T_m-T_e)/2$.
In our quantitative examples below, we 
mainly have in mind an Al$_2$O$_3$ layer of
$2\,$nm thickness, which is commonly
used as membrane or coating material in experiments
\cite{ven12,ven10,jia10,jia11,che04,liu16}. 
For such coatings, no electrochemical 
leakage is experimentally measurable 
\cite{jia10,ven12},
hence we model them as a perfect electrical 
insulator.
As a consequence, the gate electrode does
not carry an electric current 
(embedded electrode).
To avoid unrealistically high electric 
fields at the edges of electrode and membrane, 
we model them as rounded with radii
$0.5\,$nm and $1\,$nm, respectively
(see Fig. \ref{f2}).

The permittivity of the membrane
coating is denoted as $\epsilon_m$ and is
usually set to the value 
$\epsilon_m=9\,\epsilon_0$
for Al$_2$O$_3$ \cite{rob05},
where $\epsilon_0$ is the vacuum permittivity.
The surface charge density of the 
membrane coating $\sigma_m$ is caused 
in the case of Al$_2$O$_3$ by 
protonation of the hydroxyl-groups 
and is approximated as 
$\sigma_m = 0.01\,$C/m$^2$
\cite{ven12,che04,liu16,ven10}.

Likewise, for the permittivity $\epsilon_s$
and the viscosity $\eta$
we will usually choose the standard values
$\epsilon_s=80\,\epsilon_0$
and $\eta=10^{-3}\,$Pa$\cdot$s
for diluted aqueous solutions at room 
temperature.

As usual in such nanopore systems,
an external voltage difference $V$ can be 
applied between two ``external'' 
electrodes at the top and bottom 
of the system (see Fig. \ref{f1}),
giving rise to an electrical 
current through the pore.
Without loss of generality, 
the electrical potential at the
top and bottom electrodes are 
set to $V/2$ and $-V/2$, respectively.
Furthermore, a gate voltage $V_G$ 
can be applied at the gate electrode.

Finally, we assume that inside the pore
there is a cylindrical particle 
of length $L$, radius $R$, and
spherical end caps.
Of foremost interest in our present work 
will be the 
situation sketched in Fig. \ref{f2},
namely when the
cylindrical particle is oriented parallel 
to the pore axis and its center of mass
is contained in the central membrane 
plane but may be off the pore axis 
by a distance $r$.
Moreover we restrict ourselves to
perfectly stiff and relatively 
short rods.
In particular, our standard value
$L=17.2\,$nm corresponds to the case
when the particle length 
without the end caps coincides 
with the membrane thickness.

As discussed in Sect. \ref{s5}, longer DNA fragments
lead to qualitatively similar results since 
the main effects are due to the DNA segment
inside the pore, and since the persistence 
length of double stranded DNA ($50\,$nm) is 
relatively large \cite{kow12,he11,kes11}.

Situation when the particle is 
tilted relatively to the pore axis
will be briefly addressed in 
Appendix IV,
the main conclusion being that the
parallel alignment is stable 
against sufficiently small but 
otherwise arbitrary perturbations
of the orientation.

The main example we have in mind is a 
short fragment of double stranded DNA,
exhibiting under typical experimental 
conditions two electron charges per 
base pair, which are partially screened,
and thus can be modeled in terms of an 
effective radius $R=1.1\,$nm and an 
effective surface charge density 
$\sigma_p  = -0.05$ C/m$^2$ 
\cite{get13}.
Finally, the permittivity $\epsilon_p$ 
of the particle will be set to
$\epsilon_p=2\,\epsilon_0$ \cite{buy17}.

Though the cylindrical particle in 
Figs. \ref{f1}, \ref{f2} 
may also model some other 
kind of macromolecule or nanoparticle,
we henceforth often denoted 
it simply as ``DNA''.

\subsection{Theoretical framework and main observable}
\label{s22}
The main objective of our present work
is to determine the net 
force which acts in radial direction on the 
DNA in Fig. \ref{f2}.
For symmetry reasons, the direction of
this radial force is trivial (pointing along 
the connection between pore center and
rod center). Moreover, its modulus only
depends on the radial distance $r$ from 
the pore axis in  Fig. \ref{f2}
and is henceforth denoted as 
$F_r(r)$.
This force $F_r(r)$ and
the concomitant potential energy
of the DNA in radial direction
\begin{equation}
U(r):=-\int_0^r F_r(r')dr'
\label{n3}
\end{equation}
will be the quantities of foremost
interest in our subsequent discussion.

Since the equilibration of the surrounding 
medium is fast compared to the timescale 
of DNA motion, we neglect any dynamical 
back-coupling effects of the DNA motion 
on the surrounding medium, i.e., we will 
focus on steady state situations.
Furthermore, we will employ the well 
established continuum approximation
based on the coupled 
Poisson, Nernst-Planck, and Stokes 
equations (PNPS) \cite{pro03,mas06,eis96,cor00}: 
The Poisson equation reads
\begin{equation}
\nabla \left[\epsilon(\bx) \nabla \psi(\bx)\right]
= 
-\rho(\bx) \ ,
\label{n1}
\end{equation}
where $\psi(\bx)$ is the electric potential
at
$\bx:=(x_1,x_2,x_3)$,
$\epsilon(\bx)$ is the permittivity,
and $\rho(\bx)$ is the charge density due 
to the fixed surface charges and the 
mobile ion charges.

Inside the DNA (index $p$ for ``particle'')
and inside the membrane coating
(index $m$), the Poisson 
equation (\ref{n1}) reduces to 
the Laplace equation
\begin{equation}
\epsilon_k \nabla^2 \psi(\bx) = 0 \ ,
\label{n1a1}
\end{equation}
with $k\in\{p,m\}$ and where
$\epsilon_k$ are the permittivities 
from Sect. \ref{s21}.

Denoting by $c_i(\bx)$ ($i=1,2$)
the local concentrations of the two
ion species (e.g. K$^+$ and Cl$^-$),
and by $Z_{1,2}$ their valences
(usually $Z_1=1$ and $Z_2=-1$)
the total ion charge density 
amounts to
\begin{equation}
\rho(\bx)=N_A e\, [Z_1\,c_1(\bx)+Z_2\, c_2(\bx)] \ ,
\label{n1b}
\end{equation}
where $N_A$ is the Avogadro constant
and $e$ the elementary charge.
Thus, inside the electrolyte solution 
(index ``$s$''), the Poisson equation 
(\ref{n1}) takes the form
\begin{equation}
\epsilon_s \nabla^2 \psi(\bx)
= - N_A e\, [Z_1\,c_1(\bx)+Z_2\, c_2(\bx)] \ ,
\label{n1b1}
\end{equation}
with $\epsilon_s$ from Sect. \ref{s21}.

To account for the fixed surface charges, 
the so far described solutions $\psi_k(\bx)$ 
of the Poisson equation inside the 
three domains with indices $k\in\{p,m,s\}$ 
must satisfy the standard matching conditions 
\begin{equation}
\left[\epsilon_s\nabla\psi_s(\bx)
-\epsilon_k\nabla\psi_k(\bx)\right]\cdot \bn(\bx)
=-\sigma_k \ ,
\label{n1a}
\end{equation}
for all $\bx$ on the charged interfaces 
between ``$s$'' and ``$k\in\{p,m\}$'',
where $\bn(\bx)$ denotes the 
normal vector pointing 
into the electrolyte solution,
and where $\sigma_k$ is the
surface charge density of the membrane
(if $k=m$) or the DNA (if $k=p$) from
Sect. \ref{s21}.

The externally imposed potential difference 
$V$ is accounted for via boundary conditions
$\psi(\bx)|_{S_{1,2}}=\pm V/2$
on the two external electrode surfaces 
$S_1$ and $S_2$ at the top and bottom
of the container in Fig. \ref{f1}.
Likewise, the electric potential applied to 
the gate electrode is taken into account 
via the boundary condition
$\psi(\bx)|_{S_3}=V_G$ at the gate electrode 
surface $S_3$,
i.e. the interface between the yellow and 
gray regions in Fig. \ref{f2}.
Finally, we require that the normal component of 
the electrical field 
\begin{equation}
\bE(\bx):=-\nabla\psi(\bx)
\label{ef}
\end{equation}
must vanish at the cylindrical 
container walls in Fig. \ref{f1},
that is $\bE(\bx)\cdot\bn(\bx)=0$.

The Nernst Planck-equation models the
particle current $\bJ_i(\bx)$ associated 
with the $i$-th ion species ($i=1,2$),
and is given by \cite{pro03,mas06,eis96,cor00}
\begin{equation}
\bJ_{i}(\bx) = \bu(\bx)c_{i}(\bx)-D_{i}\nabla
c_{i}(\bx)-\mu_{i}c_{i}(\bx)\nabla\psi(\bx)
\ .
\label{n2}
\end{equation}
The first term on the right hand side describes 
transport due to convection, with $\bu(\bx)$ 
denoting the local fluid velocity. 
The second term is due to diffusive transport, 
involving the diffusion coefficients $D_i$
of the $i$-th ion species from Sect. \ref{s21}.
The last term describes the migration of the 
ions under the influence of the electric 
field (\ref{ef}), 
where $\mu_{i}:=Z_{i}eD_{i}/k_BT$ 
are the ion mobilities, 
$\kB$ the Boltzmann constant, and 
$T=293\,$K (room temperature).
The concomitant steady state 
continuity equations are
\begin{equation}
\nabla \bJ_{i}(\bx)=0 \ ,
\label{n1d}
\end{equation}
complemented by the following boundary conditions:
At the external electrodes 
(top and bottom of the container in Fig. \ref{f1})
the ion concentrations $c_i(\bx)$ are
required to assume the preset bulk 
value $c_0$. 
On all other fluid boundaries
(including the DNA surface)
no-flux conditions
$\bn(\bx)\cdot\bJ_{i}(\bx)=0$
are imposed.

Finally, the Navier-Stokes equation
is employed in order to determine the 
fluid flow field $\bu(\bx)$ and the 
pressure field $p(\bx)$.
Since the Reynolds number is very low
under typical experimental conditions,
we can safely neglect the inertial term,
resulting in the steady state 
Stokes equation 
\begin{equation}
\eta\, \Delta\bu(\bx)-
\nabla p(\bx)-\rho(\bx)\nabla\psi(\bx)=0 \ ,
\label{n1c}
\end{equation}
complemented by the steady state
continuity equation for 
incompressible fluids
\begin{equation}
\nabla\bu(\bx)=0 \ ,
\label{n2a}
\end{equation}
and the following boundary conditions:
At the top and bottom electrodes, 
the pressure is required to assume
a preset constant value, which
can and will be set to zero, 
since only pressure 
gradients actually matter.
Furthermore, at the top and bottom 
electrodes the shear stress 
(or normal stress)
$H(\bx)\bn(\bx)$ 
as well as the tangential fluid flow 
${\bf t}(\bx)\cdot\bu(\bx)$ must vanish,
where $H$ is the hydrodynamic stress
tensor with components 
$H_{jk}:=\eta(\partial u_j/\partial x_k
+\partial u_k/\partial x_j) -\delta_{jk}p$
(arguments $\bx$ omitted and $j,k=1,...,3$),
and where ${\bf t}(\bx)$ 
denotes an arbitrary tangential vector.
On all other fluid boundaries 
(including the DNA surface)
we impose no-slip boundary 
conditions $\bu(\bx)={\bf 0}$.

Once the above set of equations is solved,
the net force $\bf F$ acting on the
DNA is obtained as the integral
\begin{equation}
{\bf F}= \int_S [H(\bx)+M(\bx)] {\bf n}(\bx)\, dS
\label{n2b}
\end{equation}
over the DNA surface $S$.
Here, $H$ is the 
hydrodynamic stress tensor from above and
$M$ is the Maxwell stress tensor
with components
$M_{jk}:=\epsilon_s(E_jE_k-\delta_{jk}
|\bE|^2/2)$,
where $E_j$ are the components of the
electrical field from (\ref{ef}) and
arguments $\bx$ have been omitted.
Finally, the radial force component
$F_r(r)$ and the potential energy $U(r)$ from
Eq. (\ref{n3}) can be readily deduced from
the so obtained total forces $\bf F$ for various 
distances $r$ of the DNA from the pore axis.

\subsection{Approximations at thermal equilibrium}
\label{s23}
If $V=0$, the system is at thermal 
equilibrium, hence all transport
currents must vanish.
In particular,
$\bu(\bx)={\bf 0}$ and $\bJ_{i}(\bx)={\bf 0}$ 
for both ion species $i=1,2$. 
The continuity equations (\ref{n1d}) and (\ref{n2a})
are thus automatically
satisfied and from the Nernst-Planck equation
(\ref{n2}) one can infer
that the concentration fields 
$c_i(\bx)$ are Boltzmann 
distributed, 
\begin{equation}
c_i(\bx)=c_0\, \exp\left(-Z_{i} e \psi(\bx)/\kB T\right) 
\ .
\label{n4}
\end{equation}

From now on, we restrict ourselves 
to the most important case with 
valences 
\begin{eqnarray}
Z_1=-Z_2=1 \ ,
\label{n4b}
\end{eqnarray}
hence the total ion concentration amounts to
\begin{equation}
c_{tot}(\bx) :=  c_1(\bx) + c_2(\bx)
= 2 c_0 \cosh(e\psi(\bx)/\kB T)
\ .
\label{n4a}
\end{equation}
Analogously, the Stokes equation (\ref{n1c}) yields
\begin{equation}
p(\bx) = 2\kB T \NA c_0 [\cosh(e\psi(\bx)/\kB T)-1]
\ . 
\label{press}
\end{equation}
Finally, upon inserting (\ref{n4}) into
(\ref{n1b1}) we recover the 
Poisson-Boltzmann (PB) equation 
\begin{equation}
\epsilon_s \nabla^2 \psi(\bx)
= 2eN_Ac_0\sinh(e\psi(\bx)/\kB T)\ .
\label{pb}
\end{equation}

Taking for granted the approximations 
$\cosh(y)\simeq 1+y^2/2$ 
and $\sinh(y)\simeq y$,
Eq. (\ref{press}) takes
the simplified form
\begin{equation}
p(\bx) =\NA c_0 \,(e\psi(\bx))^2/\kB T
\ . 
\label{n5}
\end{equation}
and (\ref{pb}) goes over into
the Debye-H\"uckel (DH) equation
\begin{equation}
\nabla^2\psi(\bx)=\lambda_d^{-2}\psi(\bx)
\ ,
\label{dh}
\end{equation}
where $\lambda_d:=\sqrt{\epsilon_s\kB T/2 N_A c_0 e^2}$
is the characteristic screening length (Debye length).
For our usual parameter values mentioned above 
and $c_0=100\,$mM we thus obtain 
$\lambda_d\simeq 1\,$nm.

Regarding the approximations 
(\ref{n5}) and (\ref{dh}), 
two remarks are noteworthy:
(i) They will only be exploited in our 
analytical calculations in Sect. \ref{s3} 
and Appendix II.
Throughout the rest of the paper, 
we will use the exact
Eqs. (\ref{press}) and (\ref{pb}).
(ii) {\em A priori}, they seem only
justified as long as $|e\psi(\bx)|\ll \kB T$.
But in practice, they are often observed to 
still work surprisingly well for $|e\psi(\bx)|$ 
values up to several $\kB T$
(see e.g. Refs. \cite{ohs11,mas06,hun01}),
however, without a truly convincing
explanation of this fact being available
(a probable cause seems to be an accidental 
cancellation of errors).
The same situation will also be
encountered in Sect. \ref{s3} when
comparing our analytical approximations 
with numerically exact solutions.

\section{Analytical approximations}
\label{s3}

For the complex geometry from Figs. \ref{f1} 
and \ref{f2}, we did not succeed to solve
the coupled PNPS problem from Sect. \ref{s22} 
rigorously by analytical means.
To identify the main mechanisms
of the interaction between a 
charged DNA segment and an embedded electrode
we will thus reduce the complexity of the 
original, three-dimensional
setup by developing an effectively 
one-dimensional approximation which can
be tackled analytically.
In addition, we will compare this approximation
with numerical solutions.

\begin{figure}
\begin{center}
	\epsfxsize=\columnwidth
	\epsfbox{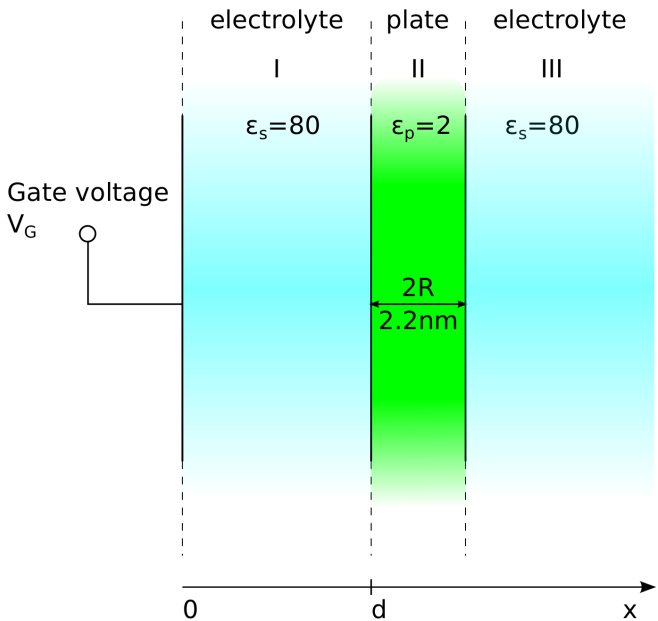} 
\end{center}
\caption{\label{f3}
(Color online)
Sketch of the simplified setup considered 
in Sect. \ref{s3}.
A planar gate electrode 
at $x=0$ is subject to a 
gate voltage $V_G$ relatively to the 
reference value $V=0$ at $x=\infty$.
The electrode is coated
by an infinitely thin, dielectric 
(non-conducting) layer with 
permittivity $\epsilon_m$ and
surface charge density $\sigma_m$
(omitted in the figure).
At a distance $d$ there
is a dielectric plate of thickness 
$2R$ parallel to the electrode
(region II, permittivity $\epsilon_p$).
The gap between electrode and 
plate (region I) as well as the region beyond 
the plate (region III) is filled with 
electrolyte solution 
(permittivity $\epsilon_s$).
}
\end{figure}

\subsection{Simplified setup}
\label{s31}
The setup considered in this section 
is sketched in Fig. \ref{f3}.
It consists of a dielectric plate 
of thickness $2R$ along the $x$-axis,
of infinite extension 
along the other two spatial directions
(perpendicular to the $x$-axis),
and with equal surface 
charge density $\sigma_p$ 
on both sides of the plate.
Furthermore, there is an infinite planar 
electrode at position $x=0$ with a preset 
electrical potential $\psi(x=0)=V_G$ relatively 
to the value $\psi(x)=0$ at $x=\infty$.
For convenience, the distance $d$
between plate and electrode 
is henceforth denoted as ``gap size''
and $V_G$ as ``gate voltage''.

The system thus splits up in three regions:
Region II represents the dielectric plate.
Regions I and III describe a
binary electrolyte solution of 
similar type as in chapter \ref{s2}
with a preset ``bulk concentration''
$c_0$ at $x=\infty$. 
Likewise, we continue to denote the 
permittivities in regions 
I, II, and III as $\epsilon_s$, 
$\epsilon_p$, and $\epsilon_s$, 
respectively. 
Most other parameters and fields
in each of the three regions
will be indicated by corresponding 
indices $j\in\{$I,II,III$\}$.

Finally, we assume that the electrode 
at $x=0$ is covered by an arbitrarily 
thin, perfectly insulating layer.
Closer inspection shows that the 
solution of the pertinent coupled 
PNPS problem (see below)
becomes independent of the actual 
permittivity $\epsilon_m$ and 
surface charge density $\sigma_m$ 
of such an infinitely thin 
dielectric layer.
The intuitive argument is that the freely
moving electrons inside the electrode can
be thought of as automatically compensating
any change in the charge distribution 
within the infinitely close-by dielectric
material. Considering that the electrode
potential $V_G$ remains unchanged, 
the overall solution of the problem 
must remain unchanged as well.
In other words, the solution of the 
problem is independent of the detailed 
charge distribution within the infinitely
thin coating and thus independent
of $\epsilon_m$ and $\sigma_m$.


\subsection{Physical interpretation and simplifications}
\label{s32}
The setup described above 
may be viewed as follows:
The embedded electrode at $x=0$ in Fig. \ref{f3}  
mimics the coated membrane electrode 
of our original setup from Fig. \ref{f2}
in the limit of an infinitely large
membrane thickness $T_m$,
an infinitely large pore radius $R_p$,
and an infinitely thin coating.
The dielectric plate in Fig. \ref{f3}  
imitates the DNA segment in Fig. \ref{f2}
in the limit of an infinite DNA 
length $L$.
Moreover, instead of just one single DNA
strand we may imagine many identical
strands, aligned in parallel and thus
approximately amounting to our 
dielectric plate of thickness 
$2R$.

It seems plausible that with respect to
the observable of foremost interest in
our present paper, namely the lateral 
or radial force $F_r(r)$
acting on the DNA in Fig.
\ref{f2} (see above Eq. (\ref{n3})),
the situation for one single DNA strand
will be different but still comparable
to the situation for many DNA strands
in parallel.
More precisely, the main observable in
Fig. \ref{f3} will be the force per 
unit area acting on the infinite plate.
The direction of that force is
obviously along the $x$ axis,
its magnitude is denoted as 
$f(d)$, and the corresponding 
potential energy $U(d)$ is defined 
similarly as in (\ref{n3}), 
namely via
$U'(d)=-f(d)$ and $U(d)\to 0$
for $d\to\infty$ 
(``pore center''),
yielding
\begin{equation}
U(d)=\int_d^\infty dx\, f(x)
\ .
\label{n10}
\end{equation}

To quantitatively compare those
forces $f(d)$ and $F_r(r)$
and the corresponding potentials 
(\ref{n3}) and (\ref{n10}), 
two quite natural further 
steps are needed:
(i) Each single DNA strand
has a diameter of $2R$ and
the relevant length inside the pore
region in Fig. \ref{f2} amounts to
$T_m$. Hence, $f(d)$ should be multiplied
by the ``effective DNA area'' 
$2RT_m$ to extract from the one-dimensional
setup in Fig. \ref{f3} a force comparable 
to $F_r(r)$.
(ii) The distance $d$ between plate 
(``DNA'') and electrode (``pore wall'')
in Fig. \ref{f3} is connected with 
the distance $r$ of the DNA from 
the pore axis in Fig. \ref{f2} 
by the relation $d=R_p-(r+R)$,
and thus
\begin{equation}
r=R_p-R-d \ .
\label{coor}
\end{equation}

In the same vein, the above mentioned
boundary conditions for the ion
concentrations and the electrical 
potential at $x=\infty$ seem reasonable
approximations to the real situation
in Fig. \ref{f2} for pores of
sufficiently large lengths $T_m$ 
and radii $R_p$.

For the rest, the transcription of the
PNPS equations
from (\ref{n1})-(\ref{n2a}) to the
present, simplified setup is straightforward 
and not explicitly carried out here.
In particular, the equations governing the
three spatial directions  
decouple from each other,
and only those in $x$-direction are 
non-trivial.
As a consequence of this decoupling property,
the solutions along the $x$-direction will be
independent from a possibly imposed
additional electrical field (or fluid flow)
orthogonal to the $x$-direction,
roughly imitating the effects of the 
externally imposed voltage $V$ in Fig. \ref{f1}.

In other words, we can and will restrict
ourselves to solutions along the
$x$-direction under thermal 
equilibrium conditions.
Since the concomitant one-dimensional 
version of the PB equation (\ref{pb}) 
still cannot be analytically solved,
we will employ the DH approximation (\ref{dh}).
Likewise, the integral (\ref{n2b})
still cannot be analytically evaluated
in terms of the exact expression 
(\ref{press}), hence we will adopt the 
approximation (\ref{n5}).
In order to validate those
two approximations, we will furthermore
compare the analytics with numerically
exact solutions in terms of  
(\ref{pb}) and (\ref{press}).

Altogether, our effectively one-dimensional 
setup from Fig. \ref{f3} is expected to 
reasonably approximate the original, three-dimensional 
setup from Figs. \ref{f1} and \ref{f2}
if the Debye length $\lambda_d$ and the 
gate voltage $V_G$ remain sufficiently 
small, and if the distance $d$ between
DNA and electrode is much
smaller than the radius $R_p$
and the length $T_m$
of the actual pore.

A somewhat similar model,
but with two infinitely 
thick, charged plates, has
been previously considered in
Ref. \cite{cha73} 
in order to study the interaction 
of two dissimilar particles
(one at constant potential and
one at constant surface charge density, 
e.g. to understand the stability 
of colloids). 
But our present setup in Fig. \ref{f3}
with a charged dielectric plate of 
{\it finite} thickness has to our 
knowledge not been worked out before.

\subsection{Qualitative expectations}
\label{s33}
For sufficiently large positive or 
negative gate potentials $V_G$ in 
Fig. \ref{f3},
it is quite plausible that the plate 
will be attracted to or repelled from 
the electrode, respectively.
Hence, there must exist a ``crossover''
value of the gate potential $V_G$, 
for which the net force acting on 
the plate is zero.
Furthermore, this crossover potential
will generically not be exactly identical 
for any given gap size $d$ in Fig. \ref{f3}
but rather will be some non-trivial 
function of $d$.
Likewise, if we consider an arbitrary but 
fixed gap size $d_0>0$ and set the gate voltage
$V_G$ to the crossover value corresponding
to $d_0$, then we expect that generically
the net force considered as a function
of $d$ changes its sign at $d=d_0$.

In other words, we predict that for suitably
chosen gate voltages $V_G$ there must exist
a gap size $d$ at which the total force
on the plate vanishes. 
The natural next question, namely 
whether this equilibrium position 
is stable or unstable against small
perturbations, cannot be decided anymore
by simple intuitive arguments.
Later on, we will see that both options
may actually be realized, depending on 
the specific values of all system
parameters.

On the qualitative level, these are 
the main results of our paper.
The main remaining goal is to quantitatively 
confirm these predictions and to
extend them to the original, 
three-dimensional setup from Figs. 
\ref{f1} and \ref{f2}.

\subsection{Analytical result and discussion}
\label{s34}
The analytical approximation announced
in Sect. \ref{s32} is worked out in 
detail in Appendix II, yielding for the
potential energy per unit area 
from (\ref{n10}) the result
\begin{eqnarray}
U(d)
& = & \frac{1}{\epsilon_s\kappa}
\left[
\frac{\gamma^2\seff^2-\sigma_p^2}{e^{2\kappa d}+\gamma^2}
+\frac{\seff\sigma_p}{\gamma\cosh(\kappa d-\ln \gamma)}
\right]
\label{1}
\\
\seff & := &  V_G\, \epsilon_s \kappa
\label{1a}
\\
\gamma & := & [1+\epsilon_p/\epsilon_s \kappa R]^{-1/2} 
\ .
\label{1b}
\end{eqnarray}
Here, $\seff$ from (\ref{1a}) may be viewed as an
effective surface charge density of the electrode 
in the absence of the plate ($d\to\infty$)
\cite{ohs11}.
As anticipated at the end of Sect. \ref{s31},
the above result does not depend on the
permittivity $\epsilon_m$ and the surface 
charge density $\sigma_m$ of the infinitely 
thin coating of the electrode.
In the limit $R\to\infty$ it follows from 
(\ref{1b}) that $\gamma\to 1$ and one 
recovers as a special case the 
result from Ref. \cite{cha73},
describing the interaction of two
infinitely thick plates, one with constant
potential and one with constant surface charge
density.

\begin{figure}[]
\begin{center}
	\epsfxsize=\columnwidth
	\epsfbox{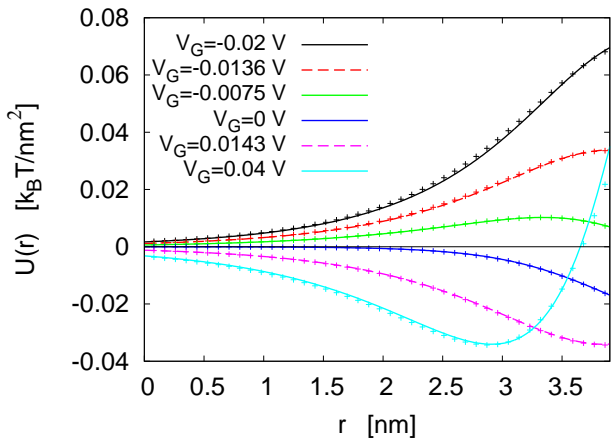} 
\end{center}
\caption{\label{f4}
(Color online)
Potential energy per area from (\ref{n10})
versus $r$ from (\ref{coor}) for the
effectively one-dimensional setup
from Fig. \ref{f3},
six different $V_G$ values as indicated,
$\sigma_p=-0.01\,$C/m$^2$, 
$c_0=100\,$mM, 
and all the remaining parameters 
as specified in Sect. \ref{s21}.
The energy is given in units of thermal
energy $k_BT$ (at room temperature) 
per surface area (in nm$^2$) of the 
dielectric plate in Fig. \ref{f3}.
Lines: Numerically exact solutions.
Symbols: Analytical approximation from 
(\ref{1}).
As detailed in the main text, the 
variable $r$ from (\ref{coor}) 
is adopted for the sake of better 
comparability with the setups 
considered later on.
Since $R=1.1\,$nm and $R_p=5\,$nm 
(see Sect. \ref{s21}), $r$ is
restricted to values 
$\leq 3.9\,$nm, and $r=3.9\,$nm
corresponds to $d=0$, i.e.
to the situation where the dielectric
plate touches the gate electrode
in Fig. \ref{f3}.
The results for $r<0$ (i.e. large $d$)
are of little interest and have been omitted.
The transition from a monotonic to a 
non-monotonic behavior of $U(d)$
takes place close to the depicted cases
with $V_G=-0.0136\,$V and
$V_G=0.0143\,$V.
}
\end{figure}

\begin{figure}
\begin{center}
	\epsfxsize=\columnwidth
	\epsfbox{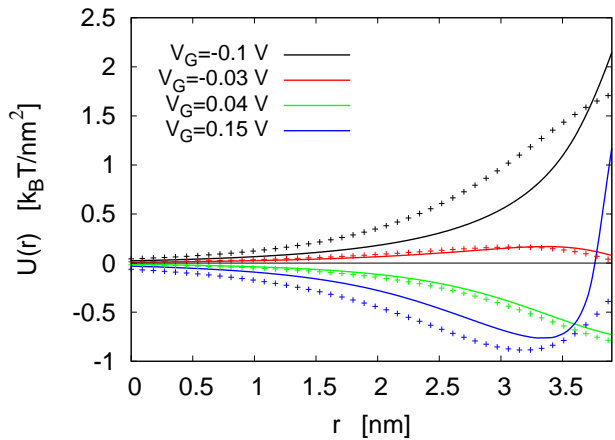} 
\end{center}
\caption{\label{f5}
(Color online)
Same as in Fig. \ref{f4} but with
$\sigma_p=-0.05\,$C/m$^2$
(i.e. the standard surface charge density
of our DNA model, see Sect. \ref{s21}) 
and modified gate voltages $V_G$.
}
\end{figure}

The approximation (\ref{1}) represents
the main analytical result of our paper.
Qualitatively, the discussion 
of its basic features is straightforward,
see Appendix III.
Depending on the quantitative values of 
the two charge densities $\seff$ and $\sigma_p$,
one finds that there are four different 
scenarios of how the function $U(d)$ may 
behave within the physically relevant domain 
$d\geq 0$:

If the signs of $\seff$ and $\sigma_p$
are equal and $|\seff| \geq |\sigma_p|$
then $U(d)$ in (\ref{1}) is monotonically decreasing.
If the signs of $\seff$ and $\sigma_p$
are equal and $|\seff| < |\sigma_p|$
then $U(d)$ exhibits a local maximum.
If the signs of $\seff$ and $\sigma_p$
are opposite and $\gamma^2|\seff| \leq |\sigma_p|$
then $U(d)$ is monotonically increasing.
If the signs of $\seff$ and $\sigma_p$
are opposite and $\gamma^2|\seff| > |\sigma_p|$
then $U(d)$ exhibits a local minimum.
Finally, if either $\seff$ or $\sigma_p$ 
is zero then $U(d)$ is 
monotonically decreasing or increasing,
respectively, and if $\seff=\sigma_p=0$
then $U(d)\equiv 0$. Typical examples, including the case $\seff=0$,
are shown in Figs. \ref{f4} and \ref{f5}.

As repeatedly mentioned, in deriving  
(\ref{1}) we have approximated
the exact Eqs. (\ref{press}) and (\ref{pb})
by (\ref{n5}) and (\ref{dh}), respectively.
To assess the validity of this approximation, 
we also included in Figs. \ref{f4} and \ref{f5} 
numerical solutions based on the exact 
Eqs. (\ref{pb}) and (\ref{press}).
As expected, when the gate voltage $V_G$ and/or
the surface charge density $\sigma_p$
increases (in modulus), the agreement between
the exact (numerical) and the approximate 
(analytical) solutions in Figs. \ref{f4} 
and \ref{f5} is seen to deteriorate.
On the other hand, and as announced at the 
end of Sect. \ref{s23}, the agreement 
still remains acceptable up to unexpectedly
large values of $|e\psi(\bx)|/\kB T$
(and thus of $|e V_G|/\kB T$).

Most importantly, the case with 
$V_G=0.15\,$V in Fig. \ref{f5}
provides a first example that 
our qualitative expectations 
from Sect. \ref{s33} may indeed
be realized, and may also
be quantitatively of interest.
Namely, this example indicates 
that a DNA with an effective surface
area of, say, $10\,$nm$^2$ can 
be trapped at a distance of about 
$1\,$nm from the ``pore wall'' 
with a trapping potential energy 
of about $8\,k_BT$, i.e.,
perturbations by thermal fluctuations 
are already reasonably weak.
Such an effective surface area
of $10\,$nm$^2$ may approximately 
arise for a DNA with radius 
$R=1.1\,$nm and a pore length of 
about $T_m=5\,$nm in Fig. \ref{f2}
(see Sect. \ref{s32}).
Also all other parameter values 
in this example from Fig. \ref{f5}
are experimentally quite 
reasonable, see also Sect. \ref{s21}.

Intuitively, the appearance of
such a potential energy minimum as
exemplified by Fig. \ref{f5}
for $V_G=0.15\,$V can be 
understood, very roughly speaking, 
as follows:
At very large distances $d$ in 
Fig. \ref{f3}, it is quite plausible
that the net force on the dielectric
plate will be zero.
Moreover, focusing on positive gate 
voltages $V_G$ and negative surface 
charge densities
$\sigma_p$ of the plate,
the electric potential $\psi(x)$ 
will be positive for $x=0$ (namely $\psi(0)=V_G$) 
and negative for $x=d$.
As $d$ decreases, the net force on the
plate will change.
Likewise, $\psi(d)$ will
change (usually monotonically increase),
while $\psi(0)=V_G$ holds
independently of $d$.
As far as the changes of the force 
on the plate are concerned,
the dominating contribution is due to the 
net pressure exerted by the electrolyte
solution on the surface at $x=d$ in 
Fig. \ref{f3}.
This claim may either appear reasonable
by itself or can be confirmed by more
detailed quantitative considerations,
at least for parameter values similar
to those in Figs. \ref{f4} 
and \ref{f5}.
This force is proportional to
the pressure in Eq. (\ref{press})
when replacing the argument $\bx$ 
by $d$.
As $d$ decreases, $\psi(d)$ increases
from its initial, negative value to
its positive limiting value $V_G$
when $d\to 0$.
Due the hyperbolic cosine
on the right hand side of (\ref{press}),
the force on the plate thus first decreases
with decreasing $d$,
reaches its minimum when $\psi(d)=0$,
and subsequently increases again.
On condition that $\psi(0)=V_G$
exceeds $|\psi(d\to\infty)|$, the resulting
total force on the plate amounts to a 
repulsion for $d\to 0$ and to an
attraction for sufficiently large $d$,
implying the existence of a potential 
minimum somewhere in between.

Analogous considerations in terms 
of the ion concentrations $c_i(d)$
from (\ref{n4}) are straightforward.
At large $d$, both concentrations
approach certain asymptotic values.
Upon decreasing $d$, one concentration
decreases and the other increases,
resulting in the quite non-trivial
behavior of the total concentration
from (\ref{n4a}): As $d$ decreases,
$c_{tot}(d)$ first decreases 
and later increases again, exactly as
for the pressure in (\ref{press})
(see also Fig. \ref{f11} and the 
discussion in Appendix I).
In other words, the behavior
of the total pressure variations are
governed by the osmotic pressure 
of the ions.

\section{Two-dimensional model}
\label{s4}
%
\begin{figure}
\begin{center}
	\epsfxsize=\columnwidth
	\epsfbox{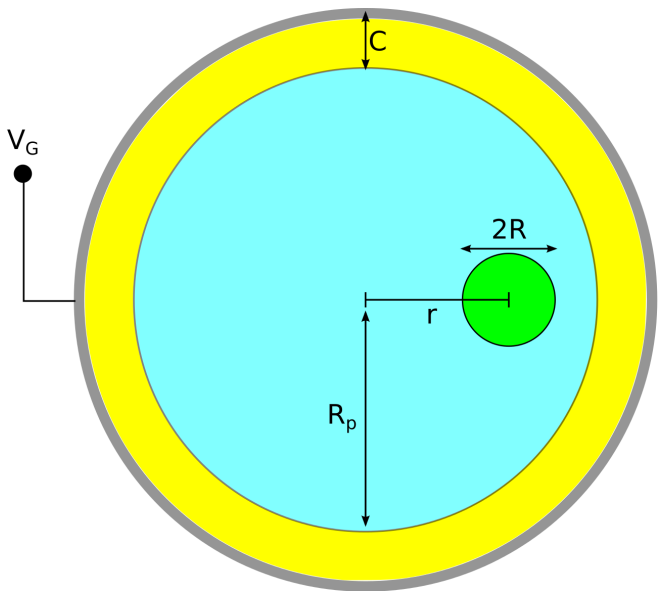} 
\end{center}
\caption{\label{f6}
(Color online)
Cross-section through the effectively
two-dimensional model, consisting of an infinitely 
long, cylindrical pore and an 
infinitely long, rod-shaped DNA
of radius $R$ parallel to the 
pore axis.
Similarly as in Figs. \ref{f1} 
and \ref{f2}, the electrolyte 
solution (blue) is confined 
by a cylindrical gate
electrode (gray) with gate
voltage $V_G$ and 
with a dielectric coating 
(yellow) of thickness $C$,
inner radius $R_p$, and surface 
charge density $\sigma_m$.
}
\end{figure}

\subsection{Description of the model}
\label{s41}
The considered setup is depicted 
in Fig. \ref{f6} and essentially
amounts to the original model
from Figs. \ref{f1} and \ref{f2} 
when the pore length $T_m$ and the 
particle length $L$ are much larger than 
the pore radius $R_p$.
In particular, the cylindrical
container and the external electrodes
in Fig. \ref{f1} are now ignored.

The further discussion is analogous
to Sect. \ref{s32}. In particular,
the general framework 
from Sect. \ref{s2} can be readily 
adapted to our present,
translation invariant setup. 
Furthermore, along the direction 
perpendicular to the plane from
Fig. \ref{f6}, the pertinent 
equations and boundary conditions
decouple from those in-plane, 
yielding a steady (or vanishing) 
motion of the fluid (electro-osmotic 
flow) and of the DNA, which we can
and will ignore from now on.
Yet another consequence of this
decoupling property is that 
the equilibrium solutions from 
Sect. \ref{s23} are applicable
for our subsequent, effectively 
two-dimensional in-plane explorations.

As before, the observable of
main interest is the force 
$f(r)$, acting in radial 
direction on the DNA per 
unit length, or equivalently,
the potential energy per
unit length
\begin{equation}
U(r):=-\int_0^r f(r')\, dr' \ .
\label{z5}
\end{equation}

\subsection{Numerical results}
\label{s43}

With the exception
of a DNA in the pore center,
which is of little interest
here, an analytical treatment
of the problem outlined in 
Sect. \ref{s41} seems impossible.
Hence we restrict ourselves to
numerical results, obtained
along the lines of 
Appendix I.

Our main focus is on the question 
to which extent the results for the 
effectively one-dimensional model from Sect. 
\ref{s3} carry over to our present, 
effectively two-dimensional model and to
the fully three-dimensional model considered in
Sect. \ref{s5}.
In a first step, we will thus 
compare the analytical approximation
from chapter \ref{s3}
with numerical solutions 
of the two-dimensional model. 
In a next step, we will identify 
particularly promising parameters 
for the fully three-dimensional explorations in 
Sect. \ref{s5}.

\begin{figure}[]
\begin{center}
	\epsfxsize=\columnwidth
	\epsfbox{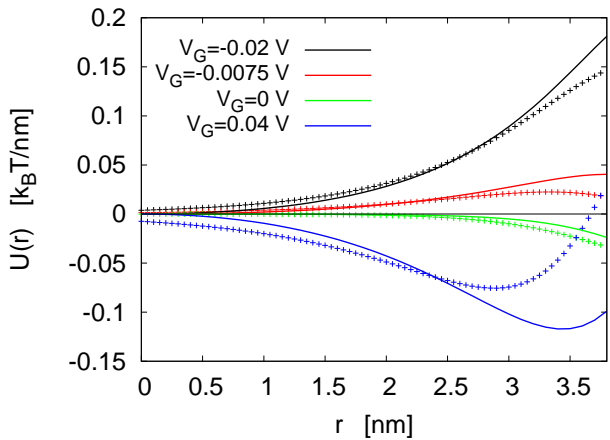} 
\end{center}
\caption{\label{f7}
(Color online)
Potential energy per unit length
versus $r$ for the effectively 
two-dimensional model from 
Sect. \ref{s41} (see also Fig. \ref{f6})
with $C=0$ (infinitely thin coating)
and pore radius $R_p=5\,$nm.
All other parameter values are as 
in Fig. \ref{f4} (except that two $V_G$ 
values have been omitted in order 
not to overload the figure).
Lines: Numerical solutions.
Symbols: Same analytical approximation
(\ref{1}) as in Fig. \ref{f4}, 
but multiplied by $2.2\,$nm to roughly
account for the ``effective DNA width'' 
in the one-dimensional model, see Sect. \ref{s32}.
}
\end{figure}

\begin{figure}
\begin{center}
	\epsfxsize=\columnwidth
	\epsfbox{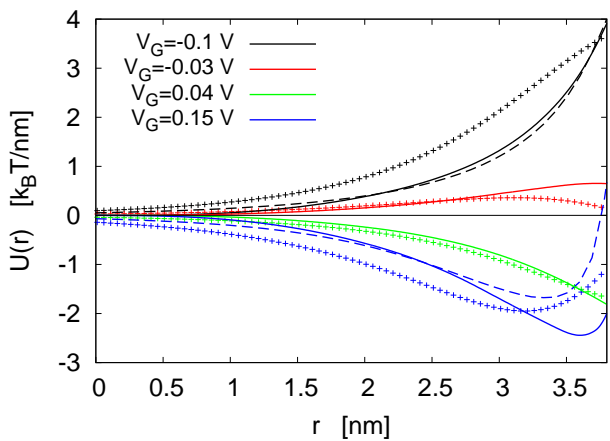} 
\end{center}
\caption{\label{f8}
(Color online)
Same as Fig. \ref{f7} except that the 
parameter values are chosen as 
in Fig. \ref{f5}.
The additional dashed lines 
for $V_G=0.15\,$V and $V_G=-0.1\,$V 
are the numerical solutions from 
Fig. \ref{f5} multiplied by $2.2\,$nm.
}
\end{figure}

For a coating of negligibly small
thickness $C$ in Fig. \ref{f6},
representative numerical results 
for the potential energy from (\ref{z5})
are depicted in Figs. \ref{f7}
and \ref{f8}, and are compared
with the analytical approximations
from Sect. \ref{s34}.
Considering that the latter approximations are
based on an effectively one-dimensional
model, the agreement is still
quite satisfying.
In particular, the existence of the
potential minima as well as their
depths and positions are reasonably
well predicted.

The dashed lines in Fig. \ref{f8} 
indicate that a considerable part 
of the deviations between the 
analytics and the numerics is not 
due to the simplified one-dimensional 
model {\em per se}, but rather to the 
additional approximations 
on which the analytics is based.
A more detailed discussion of the
main reasons for the deviations is 
beyond our present scope.

\begin{figure}
\begin{center}
	\epsfxsize=\columnwidth
	\epsfbox{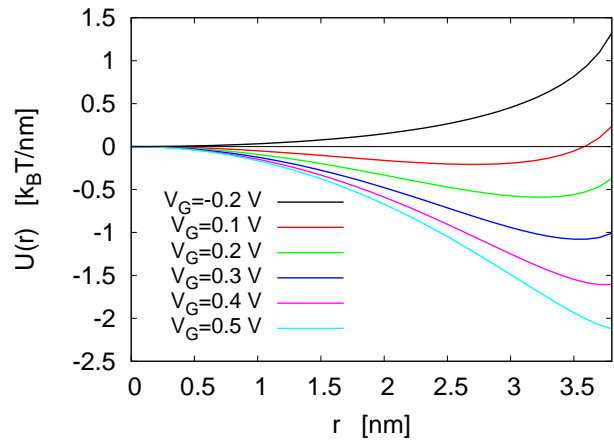} 
\end{center}
\caption{\label{f9}
(Color online)
Potential energy per unit length
versus $r$ for the effectively 
two-dimensional model from 
Sect. \ref{s41} with
pore radius $R_p=5\,$nm,
coating  thickness $C=2\,$nm 
(see Fig. \ref{f6}),
surface charge densities
$\sigma_m=0.01\,$C/m$^2$
and
$\sigma_p=-0.05\,$C/m$^2$ 
(see Sect. \ref{s21}),
and bulk concentration
$c_0=10\,$mM.
All other parameter values are as 
specified in Sect. \ref{s21}.
}
\end{figure}

Numerical results for an experimentally
more realistic coating of finite
thickness $C=2\,$nm
(see Sect. \ref{s21})
are shown in Fig. \ref{f9}.
In order to still obtain potential
energy minima with reasonable
depths and locations, a bulk 
concentration of
$c_0=10\,$mM has been adopted.
Compared to all previous numerical 
examples, the main new feature in Fig. \ref{f9} 
is that potential energies $U(r)$
which exhibit a local
maximum do not seem to exist any more.
For the rest, the qualitative behavior
is similar as before, but the quantitative 
details are very different.
In particular, analytical approximations 
are no longer available.
On the other hand, these solutions amount 
to a promising starting point for our 
later, fully three-dimensional explorations.

Analogous to Fig. \ref{f9}, we also
determined the potential energy 
landscapes for various other 
parameter values.
We desist from presenting further examples
and confine ourselves to summarizing
some of the main findings
when modifying the experimentally most
easily variable parameters:
The dependence on $V_G$ is qualitatively
similar as in Fig. \ref{f9} under
most conditions of interest.
With increasing pore radius $R_p$,
the results close to the pore walls
($r\approx R_p-R$) remain essentially 
as in Fig. \ref{f9}, but the flat regions
near $r=0$ become more extended.
For substantially smaller pore radii
$R_p$, things become more involved.

Upon increasing the coating thickness
$C$ and/or the bulk concentration 
$c_0$, the minima of $U(r)$ (within the 
region $0<r<R_p-R$) become less 
and less pronounced. The latter
tendency can be compensated to 
some extent by increasing the gate 
voltage $V_G$
(the compensation is only 
approximate, and the required 
$V_G$ values may become 
experimentally unfeasible due 
to dielectric breakdown).

As explained at the end of 
Sect. \ref{s31}, the surface charge 
density $\sigma_m$ of the coating 
material becomes irrelevant
for asymptotically thin coatings
($C\to 0$).
As $C$ grows, the dependence on
$\sigma_m$ becomes increasingly
strong, but once again, changes
of $\sigma_m$ can be (partially)
compensated by adapting the gate
voltage $V_G$.
For instance, to obtain similar
potential energy minima as in 
Fig. \ref{f9} for 
$\sigma_m=-0.05\,$C/m$^2$,
a gate voltage of about 
$V_G=1.5\,$V would be needed.
Therefore, a coating material
with a positive surface charge,
such as Al$_2$O$_3$ with 
$\sigma_m = 0.01\,$C/m$^2$
\cite{ven12,che04,liu16,ven10},
may be preferable in an actual 
experiment.

In accordance with the above observations,
it turns out that the intuitive explanation 
at the end of Sect. \ref{s34} breaks down
as the coating thickness $C$ increases.
The main reason is that the osmotic
pressure is no longer dominating the
net lateral force on the DNA.

\section{Three-dimensional model}
\label{s5}
%
%
In this Section, we present our numerical
findings for the original, fully 
three-dimensional setup from
Figs. \ref{f1} and \ref{f2}
and compare them with the effectively
two-dimensional approximation from 
Sect. \ref{s4}.

To the best of our knowledge,
comparable solutions of the
coupled PNPS problem from 
Sect. \ref{s22}, including a reasonably
sized container as in Fig. \ref{f1}
and with broken cylinder symmetry as in 
Fig. \ref{f2}, have not been previously
obtained in the context of DNA 
translocation through nanopores.
More details about the numerics
are provided in 
Appendix I.

Throughout this section, we
focus on an experimentally typical 
external voltage of $V=50\,$mV (see Fig. \ref{f1})
and on a bulk concentration of $c_0=10\,$mM.
All other parameters are as detailed
in Sect. \ref{s21}.
In particular:
pore radius $R_p=5\,$nm (see Fig. \ref{f2}),
pore length $T_m=15\,$nm,
thickness of the dielectric coating
$(T_m-T_e)/2=2\,$nm,
coating surface charge density
$\sigma_m=0.01\,$C/m$^2$,
DNA radius $R=1.1\,$nm,
DNA length $L=T_m+2R=17.2\,$nm,
and
DNA surface charge density 
$\sigma_p=-0.05\,$C/m$^2$, 
see also the drawn to scale 
Figs. \ref{f1} and \ref{f2}.

\begin{figure}
\begin{center}
	\epsfxsize=\columnwidth
	\epsfbox{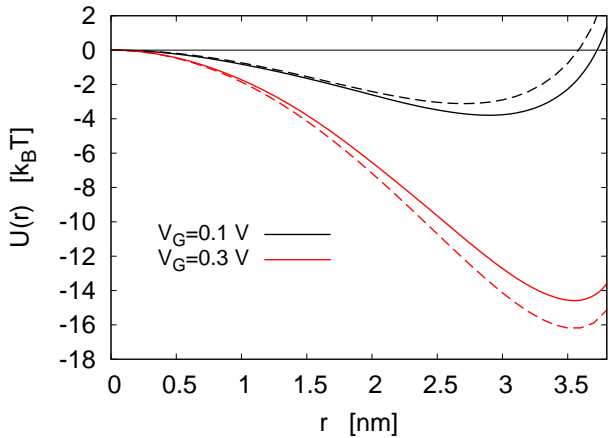} 
\end{center}
\caption{\label{f10}
(Color online) Potential
energy $U(r)$ in units of
the thermal energy $k_BT$ 
(at room temperature) for two 
different gate voltages $V_G$.
Solid: Numerical results for the
fully three-dimensional setup,
obtained from (\ref{n3}) as 
detailed in the main text.
Dashed: Same as the corresponding 
two curves in Fig. \ref{f9},
but multiplied by the pore length
$T_m=15\,$nm to approximately 
account for the ``effective system 
length'' in the two-dimensional 
model from Sect. \ref{s41}.
}
\end{figure}

Fig. \ref{f10} represents the main
result of this section.
It shows that the effectively
two-dimensional model from 
Sect. \ref{s4} approximates the
fully three-dimensional situation
very well.

In particular, the potential
energies $U(r)$ for other gate
voltages $V_G$ than shown
in Fig. \ref{f10}
behave analogously to those 
in Fig. \ref{f9}.
It follows that the range of gate 
voltages covered by Fig. \ref{f10}
is particularly interesting in 
that the potential energy $U(r)$
exhibits reasonably deep
(compared to the thermal noise 
strength $k_BT$) minima, which 
are located close (but not too 
close) to the pore wall.

A more detailed analysis of the numerical
results reveals that the two-dimensional
approximations agree remarkably well
with the three-dimensional solutions
sufficiently far inside the pore 
region in Fig. \ref{f2}
(typical differences are of the 
order of $1\%$ for $|x_2|<R_p$).
As expected, upon approaching the 
pore ends, the deviations increase 
and outside the pore the two 
solutions of course become 
entirely different.
Accordingly, the deviations between 
the solid and the dashed lines in
Fig. \ref{f10} are mainly due to 
the effects near the pore ends.
For similar reasons, the externally 
applied voltage $V$ plays a rather 
minor role.
E.g., both solid curves in 
Fig. \ref{f10} 
(obtained for $V=50\,$mV)
change (decrease) by less than 
$0.4\,k_BT$ in the case $V=0$.
Likewise, longer DNA segments
lead to quite similar results.
E.g., both solid curves in
Fig. \ref{f10}
(obtained for a DNA length
of $L=17.2\,$nm)
change (decrease) by less than
$1.5\,k_BT$ when choosing
$L=75\,$nm.
Finally, the dependence
of the lateral forces 
(and thus of $U(r)$)
on the pore length $T_m$ is,
as expected, approximately
linear (as long as $L>T_m$),
except for rather
small $T_m$.
But the latter regime is anyhow
of little interest, since  the potential
energy minima are then no longer 
sufficiently deep compared to 
the perturbations by thermal 
fluctuations.

The good agreement between the
two- and three-dimensional solutions
in Fig. \ref{f10} also implies that 
the dependence on further 
experimentally easily variable parameters 
remains approximately the same 
as discussed at the end of 
Sect. \ref{s43}.
In particular, upon increasing 
the bulk ion concentration $c_0$
one finds that
the minima of the potential 
energy $U(r)$ become less and 
less pronounced compared to 
Fig. \ref{f10}.
To some extent, this tendency can 
be compensated by reducing the
coating thickness $C$ and/or 
increasing the gate voltage 
$V_G$, but the experimental 
feasibility of both options 
may be quite limited.
The only remaining possibility
to obtain sufficiently deep
potential minima (compared 
the perturbations by thermal 
noise) is to work with 
sufficiently long pores
and particles
($L$ and $T_m$ in Fig. \ref{f2}).
Yet another consequence of increasing
$c_0$ is that appreciable variations
of $U(r)$ are confined to smaller and
smaller neighborhoods of the pore wall
(the quite obvious reason is the 
decreasing Debye screening length 
of particle and pore).
In view of the resulting very small
distances between particle and pore 
wall, one then may question the validity
of our continuum approximation
(see Sect. \ref{s22})
and of our neglecting any chemical 
and mechanical interactions between 
particle and wall (adhesion, 
friction etc.).
In conclusion, we expect that the 
most promising concentrations for 
our purposes will be relatively 
small but still doable (experiments
working with $c_0=20\,$mM are exemplified 
by Refs. \cite{spi11,dor09,gal14}).

\section{Conclusions}
\label{s6}
We have demonstrated within a simplified,
but still reasonably realistic theoretical
model that a DNA fragment in a nanopore
can be forced to traverse the pore
at a preset distance from the pore wall
by means of an embedded (dielectrically 
coated) membrane electrode, and
that this distance can be
varied by means of the applied gate voltage.

Quantitatively, Fig. \ref{f10} summarizes
our main findings, showing that the 
DNA distance from the pore wall can be
controlled up to thermal fluctuations 
of the order of $1\,$nm.
The pertinent potential energy minima 
become even more pronounced, and thus 
the DNA-wall
distance can be better controlled, as the
pore and DNA lengths are further 
increased.

Qualitatively, the existence of an extremum
of the potential energy $U(r)$ for properly
chosen system parameters can be understood
on very general grounds (see Sect. \ref{s33}).
The fact that the extremum may even be a 
minimum requires more detailed considerations,
but can still be intuitively explained
as the net effect of the osmotic pressure
contributions by the different (negatively
and positively charged) ion species
(see Sect. \ref{s34}).

Generally speaking, a passage through the 
pore which is forced to take place close 
to the pore wall is expected to exhibit 
similarities with the translocation close 
to the center of a considerably smaller pore.
In particular, the two critical issues
mentioned in the introduction, namely 
thermal fluctuations and the translocation 
speed may be considerably reduced.
More precisely, a pronounced minimum 
of the radial potential energy reduces
noise effects at least in radial 
direction and probably also with 
respect to the orientation relatively 
to the pore axis 
(see Appendix IV).
Moreover, close to the wall, the
electroosmotic drag will be reduced
and the hydrodynamic interaction with 
the wall increased, both
tending to reduce the 
translocation speed.
Finally, a reduced and better defined
distance to the pore wall will also
help to improve the resolution of 
established detection tools such as
ionic current readout \cite{dek07,bra08}, 
nanowire field-effect sensors \cite{xie12}, 
or graphene nanoribbons \cite{tra13}.
To explore the detailed quantitative 
implications for a specific setup of 
the latter type remains as an 
interesting task for future studies.

\acknowledgments
This work was supported by Deutsche Forschungsgemeinschaft
under RE1344/8-1 and RE1344/9-1, and by the Paderborn 
Center for Parallel Computing.

\section*{APPENDIX I}
In this Appendix, we provide more details
of how the numerical results form the main
text were obtained and we illustrate
how a typical numerical solution 
looks.

For the complex geometry from Figs. \ref{f1} 
and \ref{f2}, the numerical solutions of 
the coupled PNPS problem from Sect. \ref{s22} 
were obtained by means of the finite element 
software COMSOL 5.1.
The fully three-dimensional numerical
problem still requires a very high 
number (up to $2\cdot 10^7$)
of degrees of freedom, 
which is beyond of what is 
doable on normal office computers.
Therefore, we run the program on 4 cores 
with a memory requirement of up to
620\,GB RAM
at the Paderborn Center for 
Parallel Computing.

\begin{figure}
\begin{center}
	\epsfxsize=\columnwidth
	\epsfbox{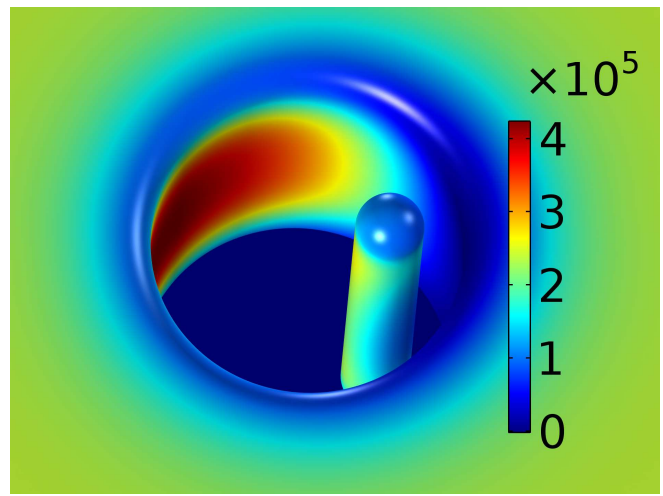} 
\end{center}
\caption{\label{f11}
(Color online) 
Representative numerical solution of the
coupled PNPS problem from Sect. \ref{s22} 
for the setup from Figs. \ref{f1}, \ref{f2} 
and with the standard
parameter values as specified in
Sect. \ref{s21}, 
external voltage
$V=50\,$mV, gate voltage 
$V_G=300\,$mV, and bulk concentration
$c_0=10\,$mM. 
Depicted is the color coded pressure (in Pa)
on the surface of the porous membrane 
and the rod shaped DNA.
More precisely, we plot the excess 
pressure over the preset value (zero) 
at the external electrodes in 
Fig. \ref{f1}.
}
\end{figure}

To illustrate the difficulty of the
problem, Fig. \ref{f11} exemplifies
the numerically obtained behavior of 
the pressure field $p(\bx)$
on the surfaces of the DNA and the membrane.
In view of these results it is quite plausible
that analytical progress will only be 
possible by means of considerable 
approximations (cf. Sect. \ref{s3}).

In fact, the results from Fig. \ref{f11}
are already quite interesting in themselves 
and anticipate some key issues of
our more detailed explorations in the main text.
Namely, one sees that the net pressure force
pushes the DNA towards the pore wall.
The main reason is as follows.
According to (\ref{n4a}) and (\ref{press}),
the pressure is governed by the total 
concentration of both ion species.
Apparently, the two dislike charged
surfaces of pore and DNA are mutually 
expelling their respective screening
ions, resulting in an overall
reduction of the total ion concentration
and hence of the pressure
between DNA and near-by pore wall.


\section*{APPENDIX II}
In this Appendix we provide 
the derivation of Eq. (\ref{1}).

As announced in Sect. \ref{s32},
we will solve the one-dimensional
version of the DH equation (\ref{dh})
in regions I and III.
Similarly, a one-dimensional Laplace equation 
like in (\ref{n1a1}) has to be solved in
the remaining region II.
Explicitly, these equations read
\begin{eqnarray}
\partial_x^2\psi_j(x)
& = &
\kappa^2\psi_j(x) \ \ \text{for j=I, III} 
\label{999}
\\
\partial_x^2\psi_j(x) 
& = & 0 
\ \ \ \ \ \ \ \ \ \ \ 
\text{for j=II} \ .
\label{10}
\end{eqnarray}   
where $\kappa:=\lambda_d^{-1}$ is the
inverse of the Debye screening length
defined below Eq. (\ref{dh}).
Their general solutions are
\begin{eqnarray}
\psi_j(x)
& = & 
A_je^{-\kappa x}+B_je^{\kappa x} 
\ \ \text{for j=I, III} 
\label{99}
\\
\psi_j(x)
& = & 
A_j+B_jx \ \ \ \ \ \ \ \ \ \ \ 
\text{for j=II}
\ ,
\label{1010}
\end{eqnarray} 
involving six integration constants $A_j$, $B_j$.
These constants are fixed by the one-dimensional
version of the matching conditions (\ref{n1a})
and the boundary conditions $\psi(x=0)=V_G$
and $\psi(x\to\infty)=0$ from Sect. \ref{s31}.

For the sake of convenience, we 
employ the dimensionless variables
\begin{eqnarray}
\hat x
& := & \kappa\,x
\label{x_c}
\\
\hat\psi(\hat x) 
& := & 
\psi(x)/\psi_0
\ ,
\label{psi_c}
\end{eqnarray}
where the reference potential
$\psi_0$ will be fixed later
(see Eq. (\ref{z4}) below).

Due to (\ref{x_c}) and (\ref{psi_c}),
Eqs. (\ref{99}) and (\ref{1010})
can be rewritten in the 
dimensionless form
\begin{eqnarray}
\hat\psi_j(\hat x)
& = &
\hat A_je^{-\hat x}+\hat B_je^{\hat x} 
\ \ \ \text{for j=I, III} 
\label{ap3}
\\
\hat\psi_j(\hat x)
& = &
\hat A_j+\hat B_j\hat x \ \ \ \ \ \ \ \ \ 
\text{for j=II}
\label{ap4}
\end{eqnarray}
with $\hat A_j:=A_j/\psi_0$
and $\hat B_j:=B_j/\psi_0$.
The corresponding one-dimensional 
boundary conditions from Sects. \ref{s22} 
and \ref{s31} take the dimensionless form
\begin{eqnarray}
\hat\psi_I(0)
& = &
V_{G}/\psi_0  
\label{9b}
\\
\hat\psi_{III}(\hat x\to\infty)
& = & 0  
\label{9a}
\\
\hat\psi_I(\hat d)
& = &
\hat\psi_{II}(\hat d) 
\label{10g}
\\
\hat\psi_{II}(\hat d+2\hat R)
& = &
\hat\psi_{III}(\hat d+2\hat R)  
\label{10a}
\\
\epsilon_s\hat\psi_I'(\hat d) 
- \epsilon_p\hat\psi_{II}'(\hat d)
& = &
\sigma_p /\kappa\psi_0 
\label{10b}
\\
\!\!\!\!\!\!\!\!\!
\epsilon_p\hat\psi_{II}'(\hat d+2\hat R) 
- \epsilon_s\hat\psi_{III}'(\hat d+2\hat R)
& = & \sigma_p /\kappa \psi_0  \ ,
\label{10c}
\end{eqnarray}
where 
\begin{eqnarray}
\hat d:=\kappa\, d\, ,\  \hat R:=\kappa\, R \ .
\label{z2}
\end{eqnarray}
Without loss of generality we can choose
\begin{equation}
\psi_0:=\sigma_p/\kappa
\label{z4}
\end{equation}
so that the right hand side
of Eqs. (\ref{10b}) and (\ref{10c})
become unity.

Later on, we will see that
for the evaluation of
the force we only 
need to determine the 
constants $\hat A_I$, $\hat B_{I}$, 
and $\hat B_{III}$.
To this end, we introduce (\ref{ap3}) 
into (\ref{9b}), (\ref{10g}), (\ref{10b}) 
and solve for $\hat A_I$ 
to obtain
\begin{equation}
\hat A_{I}(\hat d)\
= 
\frac{\frac{V_{G}}{\psi_0}e^{\hat d}
-\frac{1}{\epsilon_s}}{e^{\hat d}+\gamma^2e^{-\hat d}}
\ ,
\label{A1}
\end{equation}
where $\gamma$ is 
defined in Eq. (\ref{1b})
and where the dependence of
(\ref{A1}) on $\hat d$
is now explicitly written out
for later convenience.
Likewise, introducing (\ref{A1}) 
into (\ref{ap3}) and (\ref{9b}) 
yields
\begin{equation}
  \hat B_{I}(\hat d)= 
 \frac{\gamma^2\frac{V_{G}}{\psi_0}e^{-\hat d}
 +\frac{1}{\epsilon_s}}{e^{\hat d}+\gamma^2e^{-\hat d}}
 \ .
 \label{B1}
\end{equation}
Furthermore, Eqs. (\ref{9a}) and (\ref{ap3})
imply
\begin{equation}
 \hat B_{III}=0\ . 
\label{B1a}
\end{equation}
Finally, the dimensionful constants are
recovered as $A_{I}=\psi_0\,\hat A_{I}$
and $B_{I,III}=\psi_0\,\hat B_{I,III}$. 

Given the electric potential $\psi(\bx)$,
the remaining fields readily follow
from Eqs. (\ref{n4}), (\ref{press}),
and the text above Eq. (\ref{n4}).
Likewise, the resulting force $f(d)$ on
the plate in Fig. \ref{f3}
can be obtained analogously 
as in (\ref{n2b}),
except that $f(d)$ is now a force 
per unit area, see Sect. \ref{s32}.
Regarding the hydrodynamic stress tensor
$H_{jk}$ (see above Eq. (\ref{n2b})) and the
Maxwell stress tensor $M_{jk}$ 
(see below Eq. (\ref{n2b})), one finds that 
\begin{eqnarray}
H_{11}(x)
& = &
-p(x)
\label{a11}
\\
M_{11}(x) 
& = &
\epsilon_s\left( E_1^2(x)-E_1^2(x)/2 \right)
=
\epsilon_s\, E_1^2(x)/2
\label{m11}
\end{eqnarray}
and that -- due to the one-dimensional 
nature of the problem and the fact that 
the velocity field is zero --
all other components of 
$H_{jk}$ and $M_{jk}$ vanish.
As a consequence, only the force component
$F_1$ in (\ref{n2b}) is non-zero.
Furthermore, one finds that $F_1$ 
can be written as
\begin{eqnarray}
F_1 
& = &
[h(d)n_1(d)+h(d+2R)n_1(d+2R)]\,\Delta S \ ,
\label{a2b}
\end{eqnarray}
where $\Delta S$ denotes 
the unit surface area element and
\begin{equation}
h(x):=H_{11}(x)+M_{11}(x) \ .
\label{a2ba}
\end{equation}
Dividing (\ref{a2b}) by $\Delta S$,
taking into account (\ref{a11}) and (\ref{m11}),
and observing that $n_1(d)=-1$ and 
$n_1(d+2R)=1$ yields for the force per 
area the result
\begin{eqnarray}
f(d) &=& f_h(d) + f_{el}(d) 
\label{a3b}
\\
f_h(d) &:=&  p(d)-p(d+2R) 
\label{a4b}
\\
f_{el}(d) &:=& \frac{\epsilon_s}{2}[\psi_{III}'^2(d+2R)-\psi_{I}'^2(d)]\ .
\label{a5b}
\end{eqnarray}
In (\ref{a5b}), we exploited that 
$\psi_j'(x)=-E_1(x)$ for $j\in\{I,III\}$.

A closed analytical expression for the 
integral in (\ref{n2b}) still turns out
to be impossible when working on the right
hand side of (\ref{a4b}) with the exact 
Eq. (\ref{press}).
Therefore, we invoke its 
approximation (\ref{n5}),
as justified in more detail in Sect. \ref{s32}.
Introducing this approximation into (\ref{a4b})
and adopting the dimensionless quantities
from (\ref{x_c}), (\ref{psi_c}), and (\ref{z2}),
we can rewrite (\ref{a4b}) 
and (\ref{a5b}) as
\begin{eqnarray}
f_h(d) 
&=&  
\frac{\epsilon_s}{2} 
\kappa^2 \psi_0^2 \left({\hat\psi_I}^2(\hat d)
-{\hat\psi_{III}}^2(\hat d+2\hat R)\right) 
\label{a1c}
\\ 
f_{el}(d) 
&=& 
\frac{\epsilon_s}{2} 
\kappa^2 \psi_0^2 \left({\hat\psi_{III}'^2}(\hat d
+2\hat R)-{\hat\psi_{I}'^2}(\hat d)\right) 
\label{a2c}
\ .
\end{eqnarray}
From (\ref{B1a}) and (\ref{ap3}) we 
can conclude that
\begin{equation}
\hat\psi'_{III}(\hat x)=-\hat\psi_{III}(\hat x) 
\ .
\label{B1b}
\end{equation}
Introducing (\ref{a1c}) and (\ref{a2c})
into (\ref{a3b}) and exploiting (\ref{B1b})
yields
\begin{equation}
f(d) = \frac{\epsilon_s}{2} 
\kappa^2 \psi_0^2 \left({\hat\psi_I}^2
(\hat d)-{\hat\psi_{I}'^2}(\hat d)\right)
\label{af2}\ .
\end{equation}

By means of (\ref{ap3}), (\ref{A1}), 
and (\ref{B1}) one 
can rewrite (\ref{af2}) after 
a short calculation as
\begin{eqnarray}
f(d) &=& a_0\ {\hat A_{I}}(\hat d){\hat B_{I}}(\hat d)
\label{af0}\\
a_0 &:=& 2\epsilon_s\kappa^2\psi_0^2
\ .
\label{af}
\end{eqnarray}
Exploiting (\ref{A1}) and (\ref{B1}) 
one readily sees that
\begin{eqnarray}
{\hat A_{I}}(\hat d){\hat B_{I}}(\hat d)
&=& 
i_1(\hat d)+i_2(\hat d) 
\label{ad1}
\\
i_1(\hat d) 
&:=& 
\frac{a_1}{(e^{\hat d}+\gamma^2e^{-\hat d})^2}
\label{ad2}
\\
a_1 &:=& (\gamma V_G/\psi_0)^2-1/\epsilon_s^2
\label{c1}
\\
i_2(\hat d) 
&:=& 
a_2\,
\frac{e^{\hat d}-\gamma^2e^{-\hat d}}
{(e^{\hat d}+\gamma^2e^{-\hat d})^2}\label{ad3}
\\
a_2 
&:=& 
V_G/\epsilon_s \psi_0
\ .
\end{eqnarray}

The potential energy of the plate at 
distance $d$ is given by (\ref{n10}).
Observing (\ref{af0}) and (\ref{ad1}) it 
follows that
\begin{eqnarray}
U(d) 
&=& 
a_0\left(I_1(\hat d)+I_2(\hat d)\right) 
\label{Ud}
\\
I_1(\hat d)
&:=& 
\kappa^{-1}\int_{\hat d}^\infty d{\hat x}
\, i_1(\hat x)
\label{I1}
\\
I_2(\hat d)
&:=& 
\kappa^{-1}\int_{\hat d}^\infty d{\hat x}
\, i_2(\hat x)\label{I2}\ .
\label{an11}
\end{eqnarray}
The integral (\ref{I1}) with integrand
(\ref{ad2}) can be readily evaluated 
by substitution:
\begin{eqnarray}
I_1(\hat d)
&=& 
\frac{a_1}{2\kappa}\int_{e^{2\hat d}+\gamma^2}^\infty \frac{du}{u^2}=\frac{a_1}{2\kappa (e^{2\hat d}+\gamma^2)} 
\label{I1a}
\\
u &:=& e^{2\hat x}+\gamma^2
\ .
\end{eqnarray}
The integrand (\ref{ad3}) 
in the integral (\ref{I2}) 
can be written as
\begin{eqnarray}
I_2(\hat d)
&=& 
\frac{a_2}{\kappa}\int_{\hat d}^\infty d\hat x \frac{d}{d\hat x}\left(\frac{-1}{e^{\hat x}+\gamma^2e^{-\hat x}}\right)
\nonumber 
\\
&=& \frac{a_2}{\kappa(e^{\hat d}
+\gamma^2e^{-\hat d})}
\ .
\label{I2a}
\end{eqnarray}
Inserting (\ref{I1a}) and (\ref{I2a})
into (\ref{Ud}) and exploiting
$\sigma_e=\epsilon_s\kappa V_G$ (see (\ref{1a})) 
and (\ref{z4}) results in
\begin{eqnarray}
U(d)
=
\frac{1}{\epsilon_s\kappa}\left
(\frac{\gamma^2 \sigma_e^2-\sigma_p^2}{e^{2\hat d}+\gamma^2} 
+
\frac{2\sigma_e\sigma_p}{e^{\hat d}
+\gamma^2e^{-\hat d}}\right)
\ .
\label{Ud1}
\end{eqnarray}
Rewriting the numerator of the 
last term in (\ref{Ud1})
as $2\gamma\cosh(\hat d-\ln \gamma)$
and substituting $\hat d=\kappa d$
(see (\ref{z2})) results in 
formula (\ref{1}).

%
\section*{APPENDIX III}
In this Appendix, the statements in the 
third paragraph of Sect. \ref{s34} are 
derived.

In view of (\ref{n10}),
the extrema of the function $U(d)$
from (\ref{Ud1})
are the solutions of $f(d)=0$.
Due to (\ref{af0}) this requires 
that either $\hat A_I(\hat d)=0$ 
or $\hat B_{I}(\hat d)=0$.
Denoting the solutions of the latter
two equations as $\hat d_{max}$ and
$\hat d_{min}$, respectively,
one can infer from (\ref{A1}), (\ref{B1}) 
with (\ref{1a}), (\ref{z4}) that
\begin{eqnarray}
\hat d_{max}
& = & \ln\left(\frac{\sigma_p}{\seff}\right)
\label{dmax}
\\
\hat d_{min}
& = &\ln\left(-\frac{\gamma^2\seff}{\sigma_p}\right)
\ .
\label{dmin}
\end{eqnarray}
Closer inspection of $f'(\hat d)$ 
shows that $\hat d_{max}$ corresponds 
(as anticipated by the notation)
to a maximum of $U(d)$ and
$\hat d_{min}$ to a minimum.

Since only non-negative $d$ values
are admitted in the model from
Sect. \ref{s3}, the left hand side 
of (\ref{dmax}) only counts
as a relevant maximum if the argument of
the logarithm exceeds unity,
and likewise for (\ref{dmin}).
It follows that $U(d)$ exhibits
a maximum within the domain $d\geq 0$
if and only if $\sigma_p/\seff>0$ 
and $|\seff|<|\sigma_p|$.
Likewise, $U(d)$ exhibits
a minimum in the domain $d\geq 0$
if and only if $\sigma_p/\seff<0$ 
and
$\gamma^2|\seff|>|\sigma_p|$.
Moreover, if $\sigma_p/\seff>0$ 
and $|\seff|\geq |\sigma_p|$
then there is a maximum in the
domain $d\leq 0$ (and no further extremum),
hence $U(d)$ is monotonically 
decreasing in the domain 
$d\geq 0$.
Analogously, $U(d)$ is monotonically
increasing for $d\geq 0$ if
$\sigma_p/\seff<0$ 
and
$\gamma^2|\seff|\geq |\sigma_p|$.
These findings readily imply the 
statements in the third paragraph 
of Sect. \ref{s34}.

%
\section*{APPENDIX IV}
%
%
This Appendix is devoted to stability 
considerations of the solutions for the 
three-dimensional setup from Sect. \ref{s5}.

The motion of the idealized, rod-shaped 
and perfectly stiff DNA segment from 
Fig. \ref{f2} in radial direction
is governed by the potential energy
$U(r)$ from (\ref{n3}).
According to Fig. \ref{f10},
for gate voltages between $0.1\,$V
and $0.3\,$V, the DNA mostly remains
in the close vicinity of the 
potential minimum even in the 
presence of thermal fluctuations.
Perpendicularly to this radial 
dynamics, there will be a purely
diffusive circular motion
``around'' the pore axis.

So far we always assumed that the DNA
in Figs. \ref{f1} and \ref{f2} is 
oriented exactly parallel to the 
pore axis.
Next, we address the question how 
stable this parallel alignment is 
against small perturbations of the 
orientation.

We confine ourselves to a simple, 
qualitative argument
along the lines of the 
well-established, so-called
Derjaguin approximation \cite{mas06}:
The starting point is a DNA 
oriented parallel to the pore 
axis at a distance $r$ corresponding 
to the potential energy minimum.
Moreover, we assume that the
pore (and the DNA) is relatively
long (compared to the pore radius)
so that the effectively two-dimensional
approximation from Sect. \ref{s4}
can be safely used.
Now, we imagine that the
DNA is slightly tilted.
Except that the center of mass
must be kept fixed and the rotation
angles must be small,
this reorientation may be
chosen arbitrarily.
Next, we divide the tilted DNA 
into still reasonably long, 
equal pieces, and finally 
align each piece separately 
with the pore axis (keeping the
center of mass fixed for each 
piece).
It follows that the radial force on
each piece is still reasonably well 
described by the effectively two-dimensional 
approximation, i.e. each piece
experiences a force which tends 
to drive it into the local minimum
of the potential energy.
Moreover, the net angular momentum
on the DNA can be approximated very 
well in terms of the radial 
forces on all pieces.
The global energy minimum is
reached if and only if all pieces 
are simultaneously at the potential 
energy minimum.
Elementary geometrical considerations 
readily show that the latter is only 
possible if the original DNA
was not tilted at all.


\end{document}